# Correlation between microstructure deflections and film/substrate curvature under generalized stress fields

Massimo Camarda, Ruggero Anzalone, Giuseppe D'Arrigo, Andrea Severino, Nicolò Piluso, Andrea Canino, Francesco La Via and Antonino La Magna

Consiglio Nazionale delle Ricerche, Istituto di Microelettronica e Microsistemi CNR-IMM, Z.I. VIII Strada 5 I 95121 Catania, Italy

massimo.camarda@imm.cnr.it

**Abstract.** In this article we develop an analytical theory that correlates the macroscopic curvature of stressed film/substrate systems with the microscopic in-plane and out-of-plane deflections of planar rotators. Extending this stress-deflection relations in the case of nonlinear stress fields and validating the results with the aid of finite element simulations. We use this theory to study the heteroepitaxial growth of cubic silicon carbide on silicon (100) and discover that due, to defects generated on the silicon substrate during the carbonization process, wafer curvature techniques may not allow the determination of the stress field in the grown films either quantitatively or qualitatively.

# 1. Introduction

Conductive, dielectric, semiconducting, piezoelectric and ferroelectric thin films are extensively used for micro and nano electromechanical system (MEMS and NEMS) applications. One important parameter affecting the properties of MEMS and NEMS is the residual stress in the film, which can result in buckling, cracking, splintering and sticking problems. Presently, it is very difficult to predict the residual stress from a growth process since residual stress is strongly affected by deposition conditions and subsequent fabrication processes. Stress measurement techniques are therefore essential for both process development and process monitoring. Many suggestions for stress measurements in thin films have been made over the past several decades. The conventional method involves measuring the wafer curvature to calculate the average stress using the Stoney equation [1,2,3]. Other approaches uses the resonant frequency technique [4], nanoindentation [5], X-ray diffraction (only in the case of poly-crystals) [6], deflections of microstructures such as cantilevers [7], microbridges [8] and planar rotators[9] or bulge tests on fabricated membranes [10].

In general, a defective stress field $\sigma_{def}(z)$, i.e. the stress related to the film/substrate interface and to defects in the system, can be represented by a polynomial expansion of the type:

$$\sigma_{def}(z) = \sum_{k=0}^{\infty} \frac{\sigma_k^{def}}{k!} z^k \qquad (1)$$

where $z$ is the coordinate across the thickness and $\sigma_k^{def}$ is the k-order expansion around the $z = 0$ reference, which is generally chosen either located at the film/substrate interface or at the film's middle surface.

The Stoney equation is deduced in the uniform approximation ($\sigma_0^{def} \neq 0, \sigma_i^{def} = 0 \ \forall i > 0$):

$$\sigma_{def}(z) = \sigma_0^{def} \qquad (2)$$

and relates the macroscopic curvature of the wafer ($K_{macro}$) with the defective stress [1,2,3]:

$$K_{macro} = \frac{6 h_{film}}{h_{sub}^2} \frac{\sigma_0^{def}}{M_{sub}} \qquad (3)$$

Where $h_{sub}$ and $h_{sub}$ are the thickness of the substrate and film, respectively, $M_{sub} = E_{sub}/(1-\upsilon_{sub})$ is the biaxial elastic modulus of the substrate, with $E_{sub}$ and $\upsilon_{sub}$ being the elastic modulus and Poisson's ratio of the substrate.

On the other hand, the *residual stress*, i.e. the stress in the bowed wafer, can be determined, in the limit of the linear approximation ($\sigma_0^{res} \neq 0, \sigma_1^{res} \neq 0, \sigma_i^{res} = 0 \ \forall i > 1$) using planar rotators [11,12,13] with the following stress-displacement relations:

$$\sigma_0^{res} = \frac{L_O}{2L_C L}(M_{film}\delta_x) \tag{4}$$

$$\sigma_1^{res} = \frac{2}{L_C^2}(M_{film}\delta_z) \tag{5}$$

where $\delta_x$ is the in-plane rotation displacement, $L_C$ and $L$ are the lengths of the indicator and test beams respectively, $L_O$ is the separation between the test beams and $M_{film} = E_{film}/(1-\upsilon_{film})$ is the biaxial elastic modulus of the film. The in-plane ($\delta_x$) and out-of-plane ($\delta_z$) displacements of the indicator beam can be easily determined using optical or electron microscopes.

It is important to note that, when considering residual stress, the standard approximation used is the linear one, whereas, when considering the defective stress in the Stoney equation, the uniform approximation is generally used. Our goal is to correlate the "two" stresses, describe them within the same approximation and, finally, extend the analysis to any form of defective stress field.

# 2. Connection between macro and micro deflections and extension to nonlinear stress fields

In order to link the defective and residual stress field and to connect the micro-deflections ($\delta_x, \delta_z$) with the crystalline structure (i.e., with the defects), it is necessary to correlate the macro (Eq.3) and micro (Eq.4 and 5) stress-displacement relations. To this end, we distinguish between the "initial stress," i.e., the stress present in the system before the relaxation, and a "final stress" as the stress after the relaxation. Consequently, the stress calculated using any type of microstructure is not the stress determined using the wafer curvature technique. In fact, this technique allows for the determination of the *defective stress* $\sigma_{def}$, that is, the stress in the wafer before its bending, which is directly related to the defects in the system (see Fig. 1). On the other hand, the microstructures are fabricated on the bowed wafers; so they allow for the determination of the *residual* stress, that is, the stress after the bending. The link between the two relaxation processes is the residual stress, which represents the final stress after the wafer bowing and the initial stress of the microbending (see Fig. 1). To simplify the description we will initially consider strain fields rather than the stress, and then later convert the strain to stress using the constitutive equation of the material (assumed linear).

In the Kirchoff hypothesis [12] (i.e., lateral dimensions much greater than system thickness), valid both for the wafer and the microstructures as long as $h_{film}$ is much smaller than $L_{CL}, L_C$, and $L$, we have:

$$\varepsilon_{res}(z) = \varepsilon_{def}(z) - zK_{macro} + \varepsilon_{macro} \qquad (6)$$

$$\varepsilon_{fin}(z) = \varepsilon_{res}(z) - zK_{micro} + \varepsilon_{micro} \qquad (7)$$

the macro and micro strain ($\varepsilon_{macro}$ and $\varepsilon_{micro}$) represent the in-plane relative elongation/contraction of the entire wafer ($\varepsilon_{macro}$) or of the microstructures ($\varepsilon_{micro}$), whereas $K_{macro}$ ($K_{micro}$) is the curvature of the heterosystem (microstructures) (see Fig. 1). $\varepsilon_{def}(z)$ is the imput function of the model and is related to the defects into the system, $\varepsilon_{res}(z)$ is the strain in the system after the bowing and $\varepsilon_{fin}(z)$ is the strain in the released microstructures can be determined through µ-Raman analysis of the relaxed microstructures[14]. The link between Eqs. 5 and 6, through the residual strain $\varepsilon_{res}(z)$, is noteworthy.

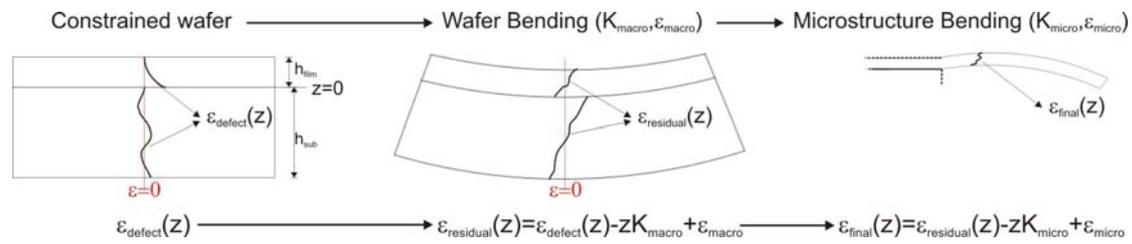

**Fig 1.** Schematic modification of the internal stress field during film growth and microstructure fabrication. a) Internal stress field of a constrained multilayered (i.e., defective or non-homogeneous) wafer. b) Residual stress field after wafer bending, this represents the initial stress condition of the fabricated microstructures (i.e., $\varepsilon_{fin}$). c) Final strain field after microstructure (and wafer) bending.

Using the connections between macro and micro deflections (Eq.6 and 7) we can generalize Eq.3-5 by minimizing the total elastic energy of the system $U$ (of the whole wafer and the microstructures) with respect to all the allowed macro and micro displacements ($K_{macro}, K_{micro}, \varepsilon_{macro}, \varepsilon_{micro}$) [15] (see Fig. 1).

$$U_{wafer}(\delta_{wafer}) = \int_{-h_{sub}}^{h_{film}} M \varepsilon_{res}^2(z, \delta_{wafer}) dz$$

$$U_{micro}(\delta_{micro}) = \int_{0}^{h_{film}} M \varepsilon_{fin}^2(z, \delta_{micro}) dz$$

where $h_{sub}$ and $h_{film}$ are the substrate and the film thicknesses respectively, with the origin of the system ($z = 0$) chosen at the substrate/film interface. $M$ is the biaxial modulus of the substrate (for $z < 0$) and film (for $z > 0$). Note the different intervals of integration for the wafer bending ($[-h_{sub}, h_{film}]$) and the microstructures' relaxation ($[0, h_{film}]$).

The micro and macro deflections can be obtained by the minimization of the total elastic energy $U$, with respect to all the allowed displacements, i.e., by setting:

$$\left.\frac{\partial U_{wafer}}{\partial K_{macro}}\right|_{\overline{K}_{macro}} = \left.\frac{\partial U_{wafer}}{\partial \varepsilon_{macro}}\right|_{\overline{\varepsilon}_{macro}} = \left.\frac{\partial U_{micro}}{\partial K_{micro}}\right|_{\overline{K}_{micro}} = \left.\frac{\partial U_{micro}}{\partial \varepsilon_{micro}}\right|_{\overline{\varepsilon}_{micro}} = 0 \quad (8)$$

In the case of macro-deflections, we have:

$$\frac{\partial U_{wafer}}{\partial K_{macro}} = -\int_{-h_{sub}}^{h_{film}} 2Mz\left(\varepsilon_{def}(z) - z\overline{K}_{macro} + \overline{\varepsilon}_{macro}\right)dz = -2\int_{-h_{sub}}^{h_{film}} Mz\varepsilon_{def}(z)dz + 2\overline{K}_{macro}\Delta_3 - 2\overline{\varepsilon}_{macro}\Delta_2$$

$$\frac{\partial U_{wafer}}{\partial \varepsilon_{macro}} = -\int_{-h_{sub}}^{h_{film}} 2M\left(\varepsilon_{def}(z) - z\overline{K}_{macro} + \overline{\varepsilon}_{macro}\right)dz = -2\int_{-h_{sub}}^{h_{film}} M\varepsilon_{def}(z)dz + 2\overline{K}_{macro}\Delta_2 - 2\overline{\varepsilon}_{macro}\Delta_1$$

From these relations, we can algebraically obtain $\overline{K}_{macro}$ and $\overline{\varepsilon}_{macro}$, i.e. the curvature and in-plane relative elongation/contraction of the system, as:

$$\overline{K}_{macro}\left(\frac{\Delta_2 \Delta_2}{\Delta_1} - \Delta_3\right) = \frac{\Delta_2}{\Delta_1}X_1 - X_1' \quad (9)$$

$$\overline{\varepsilon}_{macro}\Delta_1 = \overline{K}_{macro}\Delta_2 - X_1 \quad (10)$$

Equivalently, for the micro deflections we get:

$$\overline{K}_{micro}\left(\frac{\Theta_2 \Theta_2}{\Theta_1} - \Theta_3\right) = \frac{\Theta_2}{\Theta_1}X_2 - X_2' \quad (11)$$

$$\overline{\varepsilon}_{micro}\Theta_1 = \overline{K}_{micro}\Theta_2 - X_2 \quad (12)$$

with:

$$\Delta_1 = M_{film}h_{film} + M_{sub}h_{sub} \qquad \Theta_1 = M_{film}h_{film}$$

$$\Delta_2 = M_{film}h_{film}^2/2 - M_{sub}h_{sub}^2/2 \qquad \Theta_2 = M_{film}h_{film}^2/2$$

$$\Delta_3 = M_{film}h_{film}^3/3 + M_{sub}h_{sub}^3/3 \qquad \Theta_3 = M_{film}h_{film}^3/3$$

$$X_2 = Y_1 - \bar{K}_{macro}\Theta_2 + \bar{\varepsilon}_{macro}\Theta_1$$

$$X'_2 = Y'_1 - \bar{K}_{macro}\Theta_3 + \bar{\varepsilon}_{macro}\Theta_2$$

$$X_1 = \int_{-h_{sub}}^{h_{film}} M\varepsilon_{def}(z)dz \qquad Y_1 = \int_0^{h_{film}} M_{film}\varepsilon_{def}(z)dz$$

$$X'_1 = \int_{-h_{sub}}^{h_{film}} Mz\varepsilon_{def}(z)dz \qquad Y'_1 = \int_0^{h_{film}} M_{film}z\varepsilon_{def}(z)dz \qquad (13)$$

Eq. 9 represents the extension of the Stoney [1] equation valid for any film thickness (as in Ref. 2) and for *any* defective strain field $\varepsilon_{def}(z)$. Knowing $\varepsilon_{def}(z)$ or, equivalently, the initial defective stress field $\sigma_{def}(z) = \varepsilon_{def}(z)/M$, we can calculate the macro- and micro-deflections of the system as:

$$\bar{K}_{macro} = \left[\frac{\Delta_2}{\Delta_1}X_1 - X'_1\right] / \left(\frac{\Delta_2\Delta_2}{\Delta_1} - \Delta_3\right) \qquad (14)$$

$$\delta_z = 2L^2(\bar{K}_{macro} + \bar{K}_{micro}) \qquad (15)$$

$$\delta_x = \frac{2LL_C}{L_O}<\varepsilon_{fin}(z) - \varepsilon_{res}(z)> \qquad (16)$$

Eq. 15 has been derived noting that the out-of-plane deflection of the microstructures (either CL or PR) is considered with respect to an ideal *flat* wafer, but not with respect to the real bowed one; hence, the *measured* micro-curvature is the sum of $(\bar{K}_{macro} + \bar{K}_{micro})$. On the other hand, the test beams of the PR will elongate or contract with respect to the film/substrate stress condition (i.e., with respect to the residual stress). For this reason, the in-plane deflection, Eq. 16, is connected to the relative averaged strain $<\varepsilon_{fin}(z) - \varepsilon_{res}(z)>$ and we have:

$$\left\langle\varepsilon_{fin} - \varepsilon_{res}\right\rangle = (1/h_{film})\int_0^{h_{film}}\left(\varepsilon_{fin}(z) - \varepsilon_{res}(z)\right) = -\frac{K_{micro}h_{film}}{2} + h_{film}\varepsilon_{micro}/h_{film} =$$

$$= -\frac{K_{micro}h_{film}}{2} - \frac{K_{micro}\Theta_2}{\Theta_1} - X_2/\Theta_1 = -X_2/\Theta_1 \qquad (17)$$

So, for the in-plane deflection of the PR, we finally have:

$$\delta_x = -\frac{2LL_C}{L_O} X_2 / \Theta_1 = -\frac{2LL_C}{L_O}\left[\frac{1}{h_{film}}\int_0^{h_{film}} \varepsilon_{def}(z)dz - \frac{1}{\Theta_1}\left(\overline{K}_{macro}\Theta_2 + \overline{\varepsilon}_{macro}\Theta_1\right)\right] \quad (18)$$

where the first term on the right represents the average constrained strain of the film and the second is the modification of $\delta_x$ due to the wafer bowing.

Eq.14-16 are valid for any defective stress field, this extension can be necessary when considering stresses in crystal whose quality (i.e., the defects density) changes substantially along the thickness. This situation occurs, among the others, in the case of heteroepitaxies characterized by large lattice mismatches such as silicon carbide on silicon epitaxy (SiC/Si)[7,16,17,18,19,] or in the case of polycrystalline films where the average grain size changes, nonlinearly, along the thickness[20]. In such cases, the stress-displacement relations should be tested for extended to be valid for any possible functional form of the residual stress field. The functional form that better describes the stress field for a particular growth process can then be found by means of a careful analysis of the microstructure deflections.

We have tested this theory by comparing the analytical deflections $\delta_x$ and $\delta_z$ obtained using Eqs. 18 and 15 with those obtained using finite element simulations, assuming the same functional form of the residual stress field and the same microstructure design. To test a nonlinear functional form, we used the following exponential form for the residual stress:

$$\sigma_{def}(z) = M_{film}Ae^{-Bz} \quad z>0 \quad film$$

$$\sigma_{def}(z) = 0 \quad z<0 \quad substrate \quad (19)$$

where A and B are input parameters, to be fitted based on the observed deflections, that specify the form of the stress field along the film thickness. The analytical results have been found to be in excellent agreement with the FES, with errors below 3%, for the entire explored range of (A,B) (corresponding to a range of $\delta_x$, $\delta_z$ of $[-30,30]$ and $[-150,150]$ microns respectively, i.e., for a very large range of possible deflections).

# 3. Macro and Micro deflections for non-defective substrates

It is interesting to consider the behavior of macro- and micro-deflections as functions of the film ($h_{film}$) and substrate ($h_{sub}$) thicknesses for different possible forms of the initial constrained stress field $\sigma_{def}(z)$. This study is important for at least two reasons. The first is that by studying $\delta_x(h_{film}, h_{sub})$ and $\delta_z(h_{film}, h_{sub})$, it is possible to discern between the possible functional form for $\sigma_{def}(z)$, which is a necessary step to link $\sigma_{def}(z)$ to the defects density evolution along the film thickness and in turn to the growth parameters. The second reason is technological as when the fabricated microstructures are used for sensoring purposes, only specific maximum displacements are allowed for the microsensors to work properly. Indeed, the correct prediction of displacement as a function of thickness allows for the determination of a critical thickness above or below which the sensor will not work properly.

We have decided to test four different possible constrained fields; the uniform field,

$$\sigma_{def}(z) = M_{film} A \qquad z > 0 \tag{20}$$

which corresponds to the one used in the Stoney equations [1,2]; the linear field,

$$\sigma_{def}(z) = M_{film}(A + Bz) \qquad z > 0 \tag{21}$$

that corresponds to the one used commonly to describe residual stress fields; a power law function,

$$\sigma_{def}(z) = M_{film} A z^B \qquad z > 0 \tag{22}$$

that was found to correctly describe the evolution of stacking faults film density in the heteroepitaxial growth of SiC/Si(100)[18]; and finally, an exponentially decaying function,

$$\sigma_{def}(z) = M_{film} A e^{-Bz} \qquad z > 0 \qquad (23)$$

which was found to correctly describe the evolution of the residual stress field in the GaN/Sapphire hetero-epitaxy[21].

In all the functional forms considered, $A$ and $B$ represent input parameters, to be fitted based on the observed deflections, that describe the constrained stress field along the system thickness. In the case of uniform and linear approximation, A and B coincide with the uniform and gradient constrained strains. We can extend the concept of uniform and gradient strain/stress field in the case of nonlinear functional forms by noting that the general definition of uniform and gradient strain/stresses in the films is:

$$\sigma_{uniform} = \frac{1}{h_{film}} \int_0^{h_{film}} \sigma_{res}(z) dz \qquad (24)$$

$$\sigma_{gradient} = \frac{1}{h_{film}} \int_0^{h_{film}} \frac{\partial \sigma_{res}(z)}{\partial z} dz \qquad (25)$$

With this definition, we can calculate the uniform and gradient stresses for any functional form considered.

We have chosen only two-parameter ($A, B$) functional forms to allow for a complete determination of the stress field through the two micro-deflections ($\delta_x, \delta_z$). In general, for more complex functional forms (with more than two parameters), a sufficient number of measured deflections (varying, for example, the film thickness, as was done in Ref.[22]) are needed.

Finally, to allow for a consistent application of these different functional forms, we have chosen ($A, B$) in order to obtain the following fixed values: $\delta_z = -24 \mu m$ and $\delta_x = 3.2 \mu m$ at $h_{film} = 2.34 \mu m$ for all the considered functional forms. These specific micro-deflections have been found in the heteroepitaxial growth of 3C-SiC/Si(100) [9,23]. As can be seen in Figs. 2, 3 and 4, the four functional forms give rise to quite different results. Specifically, the out-of-plane deflection is the perfect probe to discern between linear and nonlinear constrained stress fields, as seen in Fig. 4. In the case of linear constrained fields, $\delta_z$ is independent of $h_{film}$ (and equal to zero in the uniform case). This is not what was found in the case of the heteroepitaxy of 3C-SiC/Si(100)[9,23]. This means that, in this specific growth, the constrained stress field in the film is nonlinear.

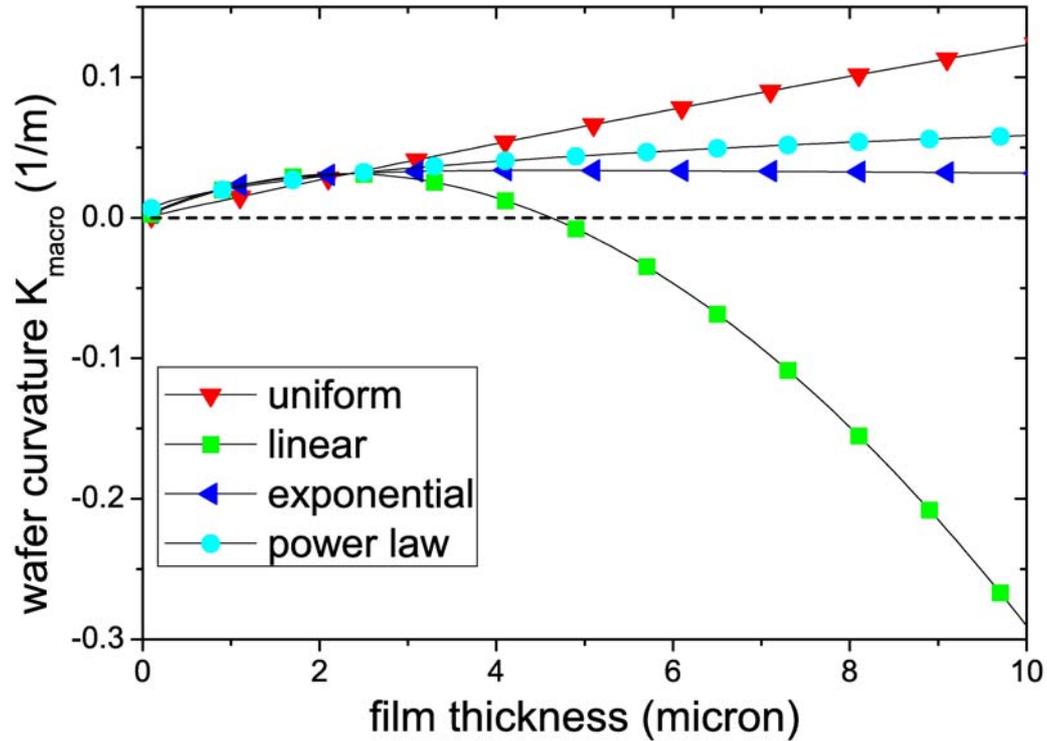

**Fig 2.** Wafer curvature as a function of film thickness in the pure uniform, linear, power law and exponential approximations. Assuming no constrained stress in the substrate. $\varepsilon_0$ and $\varepsilon_1$, for each fitting function, were chosen so that $\delta_z(h=2.34\mu m) = -24\mu m$ and $\delta_x(h=2.34\mu m) = 3.2\mu m$, consistent with the observed deflections in microstructures in the SiC/Si(100) hetero system[9,23]. The geometrical parameters of the wafer and microstructures considered are: $L_O = 20\mu m$, $L = 300\mu m$, $L_{CL} = 400\mu m$, $L_C = 160\mu m$, $w = 10\mu m$, $h_{film} = 2.34\mu m$ and $h_{sub} = 525\mu m$.

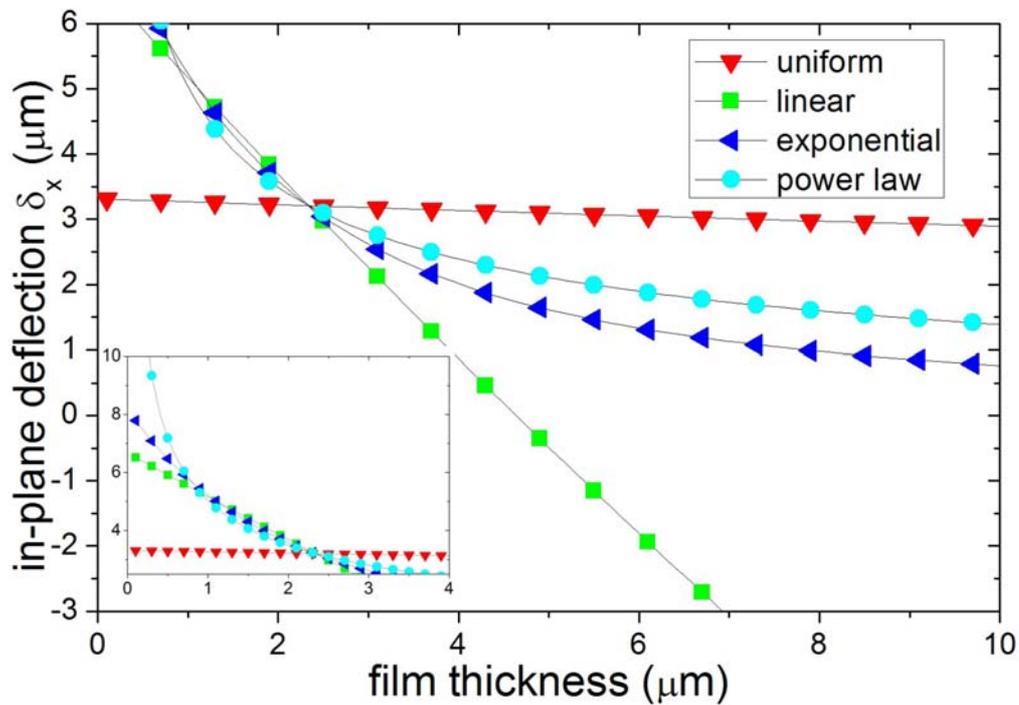

**Fig 3.** Dependency of the in-plane rotation as a function of film thickness in the pure uniform, linear, power law and exponential approximations. Inset highlights the rotation for thin films. Same parameters used as in Fig. 2

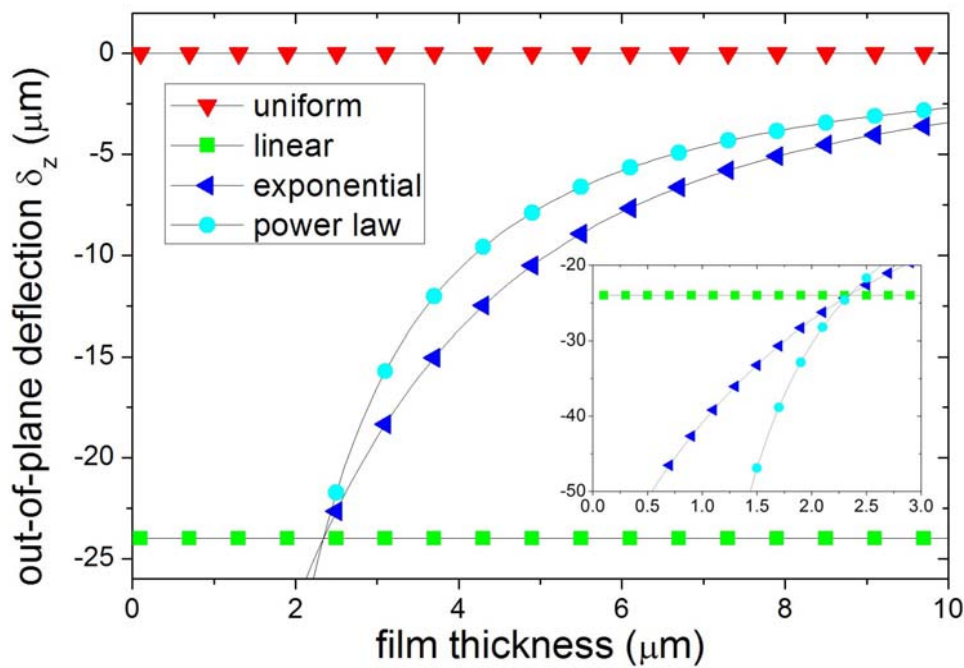

**Fig 4.** Dependency of the out-of-plane deflection as a function of film thickness in the pure uniform, linear, power law and exponential approximations. Inset highlights the deflection for thin films. Same parameters used as in Fig. 2

It is also interesting to note that in the case of linear approximation, and with the chosen parameters, there is an inversion of the wafer curvature (from concave to convex) for $h_{film} \sim 3\mu m$. This is the result of a self-compensation of the stress field inside the film. In other words, at $h_{film} \sim 3\mu m$, $\int_0^{h_{film}} \sigma_{def}(z)dz = 0$ so that we get $K_{macro} = \delta_x = 0$. This phenomenon can indeed be present in highly defective growths for which the wafer bending can be reversed for specific values of film thickness (see Fig. 2b in Ref. [24])

Note that in these conditions, the wafer can be flat even in the presence of highly stressed films. These particular conditions can be identified either by careful analysis of the out-of-plane deflections (which will be different from zero, expected in the case of an unstressed film) or through confocal $\mu$-Raman analysis that allows to investigate the residual stress field along the film thickness (see Fig. 5 and Ref. [25]).

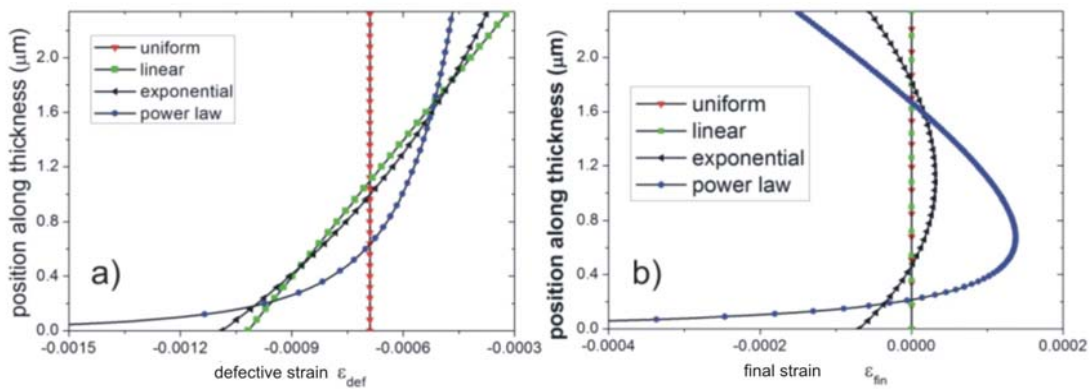

**Fig 5.** Defective, $\varepsilon_{def}(z)$, (a) and final $\varepsilon_{fin}(z)$ (b) strain fields along the film thickness. As can be seen, only nonlinear constrained fields give rise to non-zero final strains in the Kirchhoff hypothesis; on the other hand, all functional forms considered give rise to a zero final average strain $<\varepsilon_{fin}>$. Calculation parameters are the same as in Fig. 2

Finally, as can be seen in the insets of Figs. 3 and 4, the power law function has a much stronger dependency on $h_{film}$ for $h_{film} \to 0$ than the exponential function. For this reason, a careful analysis of either $\delta_x$ or $\delta_z$ for thin films ($h_{film} \sim 1\mu m$) would allow to identify the nonlinear functional form that better describes the constrained stress in the film. Finally, it is interesting to consider the initial (defective) (Eq. 6) and final stress fields (Eq. 7) in the case of the four functional forms considered. As can be seen in Fig. 5, only exponential and power law functional forms (i.e., nonlinear functional forms) give rise to nonzero final stress fields. This situation can easily be identified by confocal $\mu$-Raman analysis of the released microstructures.

## 4. Macro- and micro-deflections for defective substrates

It is important to note that for all the considered functional forms (Eqs. 20–23), we have assumed $\sigma_{def}(z) = 0$ in the substrate ($z < 0$). This assumption might not be valid for those hetero-growths where the substrate crystalline structure changes during the deposition of the film, i.e., where defects are generated in the substrate during the growth process. We will now consider this important case.

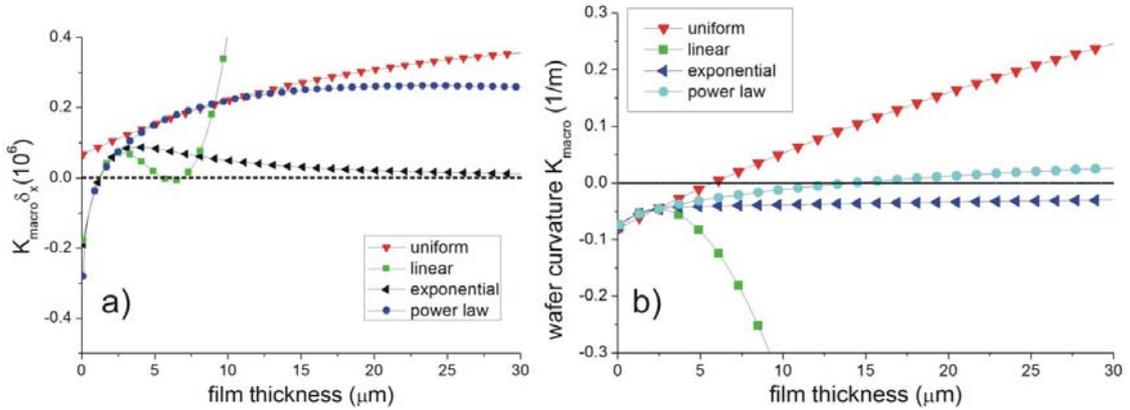

**Fig 6.** *a)* $\alpha = K_{macro}\delta_x$ and *b)* Wafer curvature, as a function of the film thickness in the case of a defective substrate ($\sigma_{def}(-h_{SubDefect} < z < 0) = -0.001$), with $h_{SubDefect} = 2\mu m$. Calculation parameters are the same as in Fig. 2. In all functional forms considered, the average film stress is tensile (except the linear form for $h_{film} > 3\mu m$, see Fig. 5)

In order to study the modifications caused by a constrained stress field located in the substrate, we start by considering the product $\alpha = K_{macro}\delta_x$ as a function of the film thickness and film stress. In the case of non-defective substrates, $\alpha$ is always greater than zero for all considered thicknesses ($h_{film}, h_{sub}$) and for all constrained functional forms (Eqs. 21–23) (see

Fig. 6a). This is related to the assumption, which is generally correct, that a tensile (compressive) stress results in a concave (convex) wafer bending [1,2,3] and in a positive (negative) PR deflection[11]. In other words, when $\sigma_{def}(z<0)=0$, the curvature of wafer and the in-plane deflection of the PR are tightly connected, and both are a direct indication of the stress status of the film (i.e., of the average constrained stress in the film; see Eq. 18). These results change when $\sigma_{def}(z<0) \neq 0$. To study this problem, we reconsidered the four constrained fields, Eqs. 20–23, and included a constant strain opposite to that present in the film and located in the substrate near the hetero-interface:

$$\varepsilon_{def}(z) = -0.0005 \qquad -h_{SubDefect} > z > 0 \mu m \qquad (26)$$

with $h_{SubDefect} = 2 \mu m$. As can be seen in Fig. 6b, in this condition, $\alpha$ can be either positive or negative depending on the film thickness. This means that, in these circumstances, a tensile (compressive) film can be associated with a convex (concave) wafer bow and that standard stress analysis based on wafer curvature analysis (Stoney equation and related [1,2,3]) would fail to predict the stress field of the film correctly. As an example, it can happen that $K_{macro} = 0$ (no wafer bow) even if the film is under a high stress condition or tensile stress and the wafer curvature is convex (see Fig. 6b). This particular situation can be identified by either Raman or XRD of the unreleased film or by analysis of the microstructure deflections. It is important to note that this situation is different from the "self-compensation" of the stress field considered in the previous section. In fact, in that case, a flat wafer ($K_{macro} = 0$) is always associated with a $\delta_x = 0$ in-plane deflection of the PR (see Fig. 3), i.e., in the case of "self-compensation", the tight link between $K_{macro}$ and $\delta_x$ is preserved. In the case of defective substrates, on the other hand, this connection is lost, so that we can have $K_{macro} = 0$ and $\delta_x \neq 0$. In such cases, only $\delta_x$, not the macroscopic bending, allows to identify the type of stress (tensile or compressive) in the epilayer correctly. It is noteworthy that in both situations, the rear of the substrate acts as an elastic constraint hindering the bowing of the system. For this reason, thicker substrates will always result in bow reduction. On the other hand, when $\int \varepsilon_{def}(z)dz = 0$, $K_{macro} = 0$ independently of the substrate thickness. This "compensated" stress condition can be important in subsequent wafer processes (backside thinning).

# 5. Analysis of the experimental results on the heteroepitaxial growth of 3C-SiC on Si(100)

Based on these considerations, we have studied the heteroepitaxial chemical vapor deposition (CVD) growth of single-crystal cubic silicon carbide (3C-SiC) on silicon (100). The growth consisted mainly of two different steps, namely, an initial "carbonization" of the silicon surface followed by the growth.

During carbonization, propane ($C_3H_8$) and hydrogen ($H_2$) flow through the reactor while the temperature is ramped to $1135°C$ at a process pressure of $\sim 400 Torr$. Once the temperature stabilizes, the wafer is held under a steady-state condition of gas flow, temperature, and pressure for four minutes. This allows the conversion of the Si wafer surface to a 3C-SiC buffer layer (carbonization).

After the four-minute carbonization plateau, the growth phase began. Silane ($SiH_4$) was then introduced into the gas stream, the temperature was ramped to $1370°C$ while the process pressure was maintained at 400 Torr. The growth rate and thickness were set equal to $2.45 \mu m/h$ and $h_{film} = 2.4 \mu m$, respectively at the wafer center. The grown film was analyzed through optical profilometer, confocal $\mu$-Raman, and microstructure deflections. As can be seen in Tab. I, there is an inconsistency between the wafer curvature analysis and the other techniques used, specifically: confocal $\mu$-Raman and PR deflections indicate a tensile residual film stress, whereas the convex wafer bow would indicate a film under compressive stress [1,2,3]. In other words, there is a clear inconsistency between wafer curvature techniques and the other analyses (Raman, and microstructure). This inconsistency, based on the results of the previous section, could be the result of a constrained stress generated in the silicon substrate during the growth process. Indeed, it is well known that during the "carbonization" step, several defects are generated underneath the interface between the 3C-SiC and the silicon substrate [26,27,28]. But these defects were not expected to play a central role in the deformation of the wafer, i.e., it was not expected that the dominant cause of the wafer bending were the defects in the substrate. In these conditions, wafer curvature analysis are not a reliable technique for the determination of the correct film stress field. To further investigate this hypothesis we have analyzed several 3C/Si samples, both on silicon (111) and (100), through cross-sectional μ-Raman to identify the stress status of the substrate throughout the entire thickness. As can be clearly seen in Fig.7, the Raman shift associated with the silicon peak increases towards higher frequencies close to the SiC/Si interface, experimentally confirming the existence of such compressed region within the substrate. Note also that the linear trend of the Raman shift (and, thus, of the strain [14,25]) far from the interface ($z < -200 \mu m$), represents the defect free region of the substrate, in fact in this region we have $\varepsilon_{res}(z) = -zK_{macro} + \varepsilon_{macro}$, being $\varepsilon_{def}(z) = 0$. It has to been noted that it was not possible to obtain such cross-section in the same epiwafer considered for the microstructures because the film was too thin to avoid de-lamination during cross-section preparation.

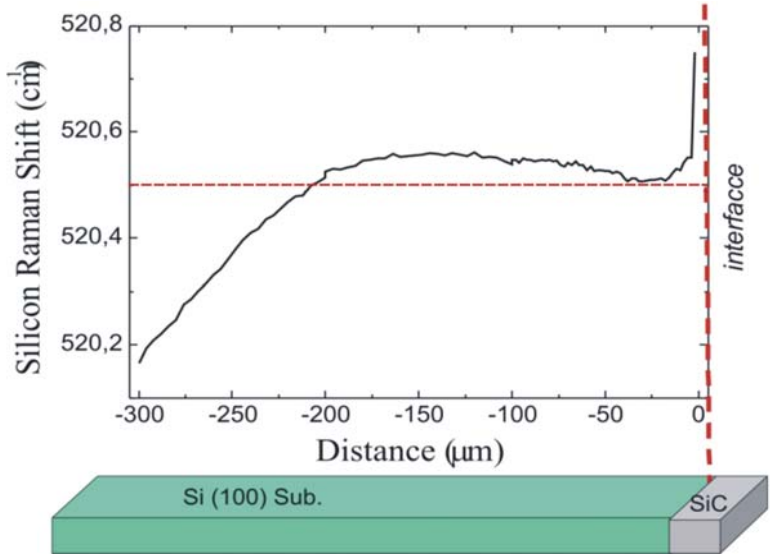

Fig.7 Cross sectional µ-Raman analysis of a 15µm SiC/Si(100) film. Red dashed line shows the stress free Raman shift value.

Finally, as a further confirmation of this interpretation, it is interesting to consider the dependency of $K_{Macro}$ on $h_{film}$ for this film. As can be seen in Fig. 6b, for $h_{film} < \bar{h}_{inv}$ (with $\bar{h}_{inv}$ depending on the chosen functional form), the curvature is a *decreasing* function of the film thickness in contrast to the usual dependency [1,2,3]. Specifically, for $h_{film} = \bar{h}_{inv}$, there is an equilibrium of the stresses, i.e., $\int_0^{h_{film}} \sigma_{def}(z)dz = \int_{-h_{sub}}^0 \sigma_{def}(z)dz$ so that we get $K_{macro} = 0$ (and $\delta_x \neq 0$). This is, indeed, what was found in Ref. [24].

Finally, it is important to note that the inversion of the curvature has been found also by varying the preparation of the substrate [29,30,31] or tuning of the carbonization step [32]. These results could be related to a different defect inclusion in the silicon substrate rather than in a change in the film microstructure. Unfortunately, the literature data do not specifically focus on the mechanical or structural analysis (e.g. determination of the Young module, strain status, or crystalline quality) of the substrate, this prevents further quantitative comparisons in the framework of the presented model.

| $h_{epi}$ [µm] | $K_{macro}$ [1/m] | LO Raman ($cm^{-1}$) | TO Raman ($cm^{-1}$) | $\delta_x$ |
|---|---|---|---|---|
| 2.34 | -0.08 | 971.58 ± 0.1 | 795.6 ± 0.1 | 3.2 ± 0.1 µm |
| | compressive[1,2,3] | tensile[10] | tensile[10] | tensile[9,11] |

**Table I:** Results of the analysis of a 3C-SiC/Si(100) epitaxially grown film. The grown film was analyzed through optical profilometer, confocal µ-Raman, and microstructure deflections. Second line shows the type of stress in the film as deduced by the analysis.

# 6. Conclusions

The aim of this work is to achieve a consistent methodology for the stress/strain analysis based on the fabrication of microstructure probes. In particular, we have developed a systematic approach to correlate global (wafer-scale) measurements to those obtained by local indicators (as cantilevers or planar rotators) of the strain status. Our interest is to push the analysis capability in case of extreme conditions of strongly nonlinear dependence of the local strain field on some material coordinate (for example, the thickness of the grown epitaxial layer). In this case, a non trivial relationship occurs between the measurable quantities (deflections and curvatures) and the strain field of the system in study.

We have derived the analytical relationships linking macroscopic and microscopic observables by formally decoupling the sequences of relaxation steps. The derived formulas have been tested in detail using numerical finite element simulations, which demonstrate the reliability of our method. The application of our formalism is useful to understand some apparent discrepancies between the measured values of macro and micro parameters in practical cases. The analysis of a particular case study on the heteroepitaxial growth of 3C-SiC on Si(100) shows that these apparent discrepancies are the signature of a complex relaxation mechanism, and an accurate data interpretation gives important insights into the microstructural configuration of the samples under study.